\documentclass[pre,floatfix,twocolumn]{revtex4}
\usepackage{subfigure} \usepackage{stdclsdv} \usepackage{array}
\usepackage{amsmath} \usepackage{amsfonts}
\usepackage{graphicx}
\usepackage{comment}

\newcommand{\avv}[1]{\langle #1\rangle}

\usepackage{color} 

\begin{document}

\title{ From Liquid Structure to Configurational Entropy: Introducing
  Structural Covariance }

\author{Pierre Ronceray} \affiliation{LPTMS, CNRS, Univ. Paris-Sud,
Universit\'e Paris-Saclay, 91405 Orsay, France \\}

\author{and Peter Harrowell} \affiliation{School of Chemistry,
University of Sydney, Sydney N.S.W. 2006, Australia}

\begin{abstract}
  We connect the configurational entropy of a liquid to the
  geometrical properties of its local energy landscape, using a
  high-temperature expansion.  It is proposed that correlations
  between local structures arises from their overlap and, being
  geometrical in nature, can be usefully determined using the inherent
  structures of high temperature liquids. We show quantitatively how
  the high-temperature covariance of these local structural
  fluctuations arising from their geometrical overlap, combined with
  their energetic stability, control the decrease of entropy with
  decreasing energy.  We apply this formalism to a family of Favoured
  Local Structure (FLS) lattice models with two low symmetry FLS's
  which are found to either crystallize or form a glass on
  cooling. The covariance, crystal energy and estimated freezing
  temperature are tested as possible predictors of glass-forming
  ability in the model system.
\end{abstract}

\maketitle

\section{ Introduction}

Why study liquid structure? The longstanding answer to this question
is that structure is expected to provide a microscopic account of the
experimentally measured properties of liquids. The theoretical
treatment of these properties,\emph{ e.g.} the equation of state and
the structure factor, are the goals that drove the development of
liquid state theory through the 1960's and
70's~\cite{liquid-theory}. The structural information required for
these tasks turned out to be modest.  The scattering function only
requires the distribution of pairwise separations and, if we can
assume pairwise additive interaction potentials, so too does the
equation of state~\cite{liquid-theory}.  Closures of correlation
function hierarchies have provided reasonable radial distribution
functions (at least for single component atomic liquids) without
explicit reference to the geometrical structure of the
liquid~\cite{liquid-theory}. This bypassing of most of the liquid
structure in the development of liquid theory is one of cornerstones
of its successes. It does leave us, however, still casting around for
a motivation to study the full many-body aspects of liquid structure.

A second reason for the study of liquid structure is related to the
properties of deeply supercooled liquids. The slow relaxation of
viscous liquids~\cite{dynamics} and the rigidity of amorphous
solids~\cite{rigidity} arise as a consequence of the constraint that a
configuration of the system exerts on the motion of its constituent
particles at low temperature. This second goal of the study of liquid
structure is, therefore, to uncover an intelligible rationale for the
origin and persistence of these constraints.  How structure might
accomplish this goal is worth fleshing out.  A structural description,
here taken to mean a reduced description of a configuration in terms
of some set of \emph{local structures} -- \emph{e.g.} the coordination
polyhedra -- is just an exercise in descriptive geometry, with the
choice of structural components left to the whim of the researcher. A
geometrical analysis of a configuration of the system allows, for each
of its constituent particles, to identify the specific local structure
that characterizes its molecular neighbourhood among a dictionary of
possible local environments.  The utility of a particular structural
description can be assessed by the information compression that it
provides. The connection to physics arises because this information
compression relies on the presence of geometrical constraints. A
particular coordination polyhedron, for example, embodies a number of
constraints between bond angles, numbers of nearest neighbours,
topological connectivity, \emph{etc.} It is these geometrical
constraints that permit the information reduction in the first
place. The hope that underlies this structural analysis is that these
\emph{geometrical} constraints on structure translate into {\it
  physical} constraints on particle positions.  A well studied example
of the fulfillment of this hope is a binary Lennard-Jones mixture
introduced by Wahnstr\"{o}m~\cite{whan} for which the correlation
between local icosahedral topology and slow relaxation has been well
established~\cite{whan-dyn}. This \emph{structure} $\rightarrow$
\emph{constraint} program depends on the information reduction
achieved by a given structural analysis. Too complex a liquid
structure and the associated information compression is insufficient
to result in any significant physical constraints. Many supercooled
liquids exhibit structures best characterised by as a diverse range of
local structures, all occurring with comparable but low frequency. In
these cases, it is difficult to see how broadly diverse structure can
be linked to the presence of physical constraint.

In this paper we concentrate on a third motivation for studying liquid
structure, one whose utility does not depend on whether the liquid
structure is simple or complex and that has received little attention
in the literature. We explore how the liquid structure connects to the
configurational entropy of the liquid and, hence, plays a crucial role
in determining the stability of the liquid with respect to
crystallization.

The concept of a configurational entropy plays a central role in the
modern description of supercooled liquids~\cite{config}. The proposal
is that the rough energy landscape that characterizes a
supercooled liquid can be usefully described in terms of local minima
of the potential energy -- the \emph{inherent structures} of the
system -- and fluctuations around these minima. The total entropy of
the liquid can thus be resolved into two parts -- the first one,
termed \emph{configurational entropy}, associated with the choice of
the inherent structure, and the second with the fluctuations of
particles around these metastable configurations.  Computationally,
the configurational entropy has been defined explicitly as $S_c =
k_{B}\log N_\mathrm{is}$ where $N_\mathrm{is}$ is the number of
distinct inherent structure~\cite{config-comp}. Calculations based on
this definition of the configurational entropy have been carried out
for binary Lennard-Jones mixtures~\cite{config-mix},
silica~\cite{config-silica} along with other model
liquids~\cite{config-other}.  Such calculations reveal that over a
wide range of temperatures, the configurational entropy can be
reasonably described as a quadratic function of the potential energy:
\begin{equation}
  S_c = S_{\infty}-\frac{1}{2K} (E-E_{\infty})^{2}
\label{eq:conf-sim}
\end{equation}
The quantities $S_{\infty}$ and $E_\infty$ respectively correspond to
the entropy and energy of inherent structures in the high temperature
limit. This expression was shown to provide a good approximation of
the configurational entropy for both a binary atomic
mixture~\cite{entropy-mix} and a model of
silica~\cite{entropy-silica}.  Sastry, Debenedetti and
Stillinger~\cite{stillinger} have demonstrated the existence of this
high-temperature limit by showing that there exists a temperature,
$T_{x}$, above which the average energy of inherent structures does
not depend on the temperature of the initial liquid. The idea is that
with sufficient thermal energy, the liquid energy lies above the
``top'' of the potential energy landscape and so is able to
effectively sample all possible inherent structures.   Referring to the empirical expression in Equation~\ref{eq:conf-sim}, the change of liquid's thermodynamic quantities upon
cooling are then controlled by the quantity $K$, which determines the
entropic cost of lowering the energy.

The aim of this article is to relate the thermodynamic properties of a
liquid -- and in particular the value of $K$ -- to its microscopic
structural origins.  The liquid's local structural properties will be
described macroscopically in terms of the frequencies of each type of
local structure. Local structures couple directly to the energy of the
system: indeed, not all local environments around a particle are
equivalent from an energetic point of view. For a given Hamiltonian,
some local structures are particularly stable, while others are
disfavoured; the energies associated to local structures constitute
the \emph{local energy landscape} of the system. On cooling, the
liquid energy will decrease, which implies that the frequencies of
low-energy local structures will increase. On the other hand, this
accumulation of stable local structures comes with an entropy cost:
indeed, the structural frequencies constrain the associated size of
the configuration space, and thus the configurational entropy. The
geometrical properties of the local structures have a direct influence
on this configurational entropy, and thus on the thermodynamic
properties of the system.

In this article, we present an analysis of liquid structure that goes
beyond mere description, and allows quantitative predictions that
become exact at high temperature, while retaining relevant information
at low temperature. We focus on the role of correlations between the
frequencies of local structures arising from their geometrical overlap
-- the \emph{structural covariances}. We show their crucial importance
in connecting the local energy landscape -- \emph{i.e.} the energy of
local structures -- to the global energy landscape -- characterized by
the configurational entropy of the system.  In
Section~\ref{sec:theory}, we present our theoretical framework in a
general context of off-lattice liquids, establishing a relationship
between liquid structure and configurational entropy.  While we expect
our predictions to hold in such cases, we have so far only tested them
in the context of a class of lattice models of liquids inspired from
our previous works, the Favoured Local Structures model with two
stable local structures, as presented in Section~\ref{sec:fls}.  Our
goals in this Section are to provide some measure of the validity of
the high temperature approximation. Employing our structural
covariance framework, we present new results concerning a version of
the FLS lattice model that appears to capture important aspects of the
problem of glass-forming ability.

\section{Local Liquid Structures and their Covariance} \label{sec:theory}

In this Section, we develop a theoretical relation between local
structure and configurational entropy in an off-lattice liquid.  There
is some assumed degree of arbitrariness in the definition of local
structures, as already discussed in the Introduction, and the precise
definition and choice of local structural parameter is usually adapted
to the specific needs of the study. A number of choices for these
structural components have been considered in the literature --
Voronoi polyhedra~\cite{bernal}, Delaunay tessellations~\cite{ogawa},
common neighbours~\cite{jonsson}, bond ring statistics~\cite{snook},
bond orientation order~\cite{steinhardt} and
tetrahedrality~\cite{doye}. One of the most common is the local
coordination polyhedra~\cite{paddy} – similar to a Delaunay vertex
except that the neighbours are defined by a cutoff separation rather
than a geometric condition based on space-filling. The literature
regarding these various choices has been summarised in a number of
excellent reviews~\cite{paddy}. The choice of structural components
depends on the goal of the analysis and the expectation of the type of
structure that might be found.  Voronoi or Delaunay tilings address
the question of how structure fills space. Common neighbour analysis
represents a reduction to the smallest non-trivial set of structural
components out of which distinct collective structures can be
assembled. Tetrahedrality focuses on the structural significance of
triangulation with respect to mechanical stability. It is important to
remark that for all definitions of local structure, there is some
degree of overlap between a local structure and its neighbours. For
instance, Voronoi polyhedra share faces, while a particle contributes
to the local coordination polyhedra of all its neighbours. This
implies that the local structure at a point imposes geometrical
constraints on the surrounding local structures.  The overlap of
coordination polyhedra was the basis of the geometrical construction
used by Frank and Kasper in their classic papers on tetrahedral
solids~\cite{kasper}.

In the analysis presented below, we consider only the \emph{inherent
  structures} of a liquid. These are the configurations associated
with local minima of the potential energy. We assume that in such
states, the environment of each particle is well characterized by its
local structure, and that the total potential energy can be written as
a sum of the energies of the individual local structures.  This
assumption implies that the system can be fully described in terms of
its \emph{local energy landscape}, \emph{i.e.} the local energy
associated with possible structural components.  The coordination
polyhedra about each particle (in concert with an assumed short range
interaction) represents a good realization of these requirements and,
for concreteness, we will select it as our definition for local
structural components in the following.

\subsection{The Structure and Energy Vectors}

Having chosen a definition of local structures, we can establish a
dictionary of all possibilities for such local structures, which we
assume to come in finite number. Implicit in this finite list of "possible" structures is that we can solve the non-trivial problem of identifying all dense local packings that avoid prohibitive overlap of particles. In what follows we shall assume that this problem can be addressed by simulation. An individual inherent structure of
the liquid can be characterized by the frequency of each type of local
structure. This introduces the \emph{structure vector} $\vec{c}$ of a
state of the system, where $c_i$ is the number concentration of
particles characterized by the \emph{i}th local structure in that
state. This vector must obey $\sum_i c_i = 1$, as we assume to
describe the full set of possible structures -- or, at least, all
structures that are likely to be observed in practice. The average
structure vector $\langle\vec{c}\rangle$ over inherent structures for
a given liquid corresponds to the typical data reported from
structural studies of simulated liquids. The pioneering 1977 studies
of Tanemura \emph{et al.}~\cite{hiwatari}, for example, consisted of
reporting the average concentrations $\langle\vec{c}\rangle$ of
Voronoi polyhedra in a soft sphere liquid.

Our key assumption, which should be valid as long as the interactions
are short ranged, is that the sole data of structure concentrations,
\emph{i.e.} the structure vector $\vec{c}$, allows for a reasonable
estimate of the energy of inherent structures. This means that the
potential energy per particle of a given inherent structure of the
system can be written simply as
\begin{equation}
  E = \vec{c}\cdot \vec{\epsilon}
\label{eq:energy}
\end{equation}
where $\vec{\epsilon}$ is the \emph{energy vector} characterizing the
local energy landscape of the system, \emph{i.e.} a site with local
structure $i$ contributes with an energy $\epsilon_i$ to the
Hamiltonian.  A simple choice for a vector with elements $\epsilon_i$
corresponding to half the average energy of a particle at the center
of the in the \emph{i}th coordination polyhedron. Since both total
energy and the $\epsilon_{i}$’s can be directly calculated, the
accuracy of Eq.~\ref{eq:energy} can be tested for each specific
system. Clearly this representation of the system is a simplification,
as it does not include, for instance, the energy associated with
distortions of local structures. From the point of view of
Eq.~\ref{eq:energy}, the total energy thus consists in a sum over all
local structures, as previously used by Procaccia and
coworkers~\cite{procaccia} in a treatment of supercooled liquids based
on the resolution of the structure into quasi-species.  There is
therefore no energetic coupling between neighbouring local structures:
the interactions are mediated solely by the geometrical constraints
arising from their overlap. This means that the identity of the
coordination polyhedron of a particle will exercise a significant
geometrical constraint on the possible coordination polyhedra of its
neighbours.  Such constraints have a purely geometrical origin, and
are thus athermal.

\subsection{ Configurational Entropy in Structure Space}

On cooling, the most stable local structures -- those with low
$\epsilon_i$'s -- are expected to accumulate in the liquid.  This
energy decrease must be balanced against the associated loss of
entropy incurred by the accumulation of the stable subset of local
structures. In this Section, we relate this entropy cost to the
geometrical properties of these stable local structures.  We have
shown recently that the accumulation of high-symmetry structures is
entropically penalized, as compared with low-symmetry
structures~\cite{ronceray2}. We now go further and encompass
correlations between structures in our description.

In the high temperature limit $T>T_x$ where the average energy of
inherent structures become temperature-independent, it is natural to
assume that their structure also reaches a limiting composition, which
we denote by $ \vec{c}_\infty = \avv{ \vec{c}}_{T=\infty}$. We are
however unaware of any examination of this proposal and whether in
fact, the temperature independence of the average inherent structure
energy is indeed translated into a temperature independent set of high
T inherent structure. A true high temperature structure, under a
constraint of fixed density, should correspond to the structure of
purely repulsive particles, a physical limit in which the purely
entropic character of the correlations become explicit.

We now consider the space formed by the possible values of $\vec{c}$
for a macroscopic system. To each such structure vector we can
associate a configurational entropy
\begin{equation}
S_c(\vec{c})=\log\Omega(\vec{c})
\label{eq:entropy-structurespace}
\end{equation}
where $\Omega(\vec{c})$ is the number of states of the liquid
compatible with the macroscopic structural composition $\vec{c}$. By
construction $S_c(\vec{c})$ is maximal at the infinite-temperature
composition $\vec{c}_\infty$, and to second order in
$\vec{c}-\vec{c}_\infty$ we can thus write:
\begin{equation}
 S_c(\vec{c})=S_{\infty}-\frac{1}{2} (\vec{c} - \vec{c}_\infty)^{T}\cdot \hat{C}^{-1} \cdot (\vec{c} - \vec{c}_\infty) + O(|\vec{c} - \vec{c}_\infty|^3)
\label{eq:entropy}
\end{equation}
where $\hat{C}^{-1}$ is a positive definite matrix characteristic of
the structure space of the liquid. This implies that the density of
states in the structure space is approximated by a multi-dimensional
Gaussian distribution,
\begin{equation}
  \Omega(\vec{c}) \approx \Omega _{\infty} \exp\left(-\frac{N}{2}(\vec{c} - \vec{c}_\infty)^{T} \cdot \hat{C}^{-1} \cdot (\vec{c} - \vec{c}_\infty)\right)
\label{eq:omega}
\end{equation}
where $N$ is the number of particles in the system. The coefficients
of the matrix $\hat{C}$ thus correspond to the high-temperature
covariances of the local structure concentrations, \emph{i.e.}
\begin{equation}
  \label{eq:C}
  C_{ij} = N\ \mathrm{Cov}_{T=\infty}(c_{i},c_{j}) 
\end{equation}
\label{eq:covar}
In the infinite-temperature limit, structural correlations become
short-ranged, and only the athermal constraints originating from the
overlap of neighbouring structures are expected to contribute to the
coefficients $C_{ij}$. While this property is exact in lattice models
as considered in Section~\ref{sec:fls}, it remains to be established
in the case of off-lattice inherent structures.

\subsection{Thermodynamic Implications of Structural Covariance }

The connection between the structure space and the more usual
microcanonical ensemble configurational entropy $S_c(E)$ is obtained
by using Eq.~\ref{eq:energy}, and integrating over structural
composition to get the density of states $\Omega(E)$ as a function of
the energy per particle $E$:
\begin{equation}
  \label{eq:omega-E-interm}
  \Omega(E) = \int \delta( E - \vec{c} \cdot \vec{\epsilon} )  \ \Omega(\vec{c})\  \mathrm{d}\vec{c}
\end{equation}
which, employing the Gaussian approximation for the density of states
introduced in Eq.~\ref{eq:omega}, yields:
\begin{equation}
  \label{eq:omega-E}
  \Omega(E) = \Omega_\infty \exp \left( \frac{(E-E_\infty)^2}{2 K}  \right)
\end{equation}
 where
\begin{equation}
  K = \vec{\epsilon}^{T} \cdot \hat{C} \cdot  \vec{\epsilon}\label{eq:K}
\end{equation}
Note that the unit vector, $(1,\dots,1)$, is an eigenvector with zero
eigenvalue of $\hat{C}$ (as departing from $\sum_i c_i = 1$ is
impossible, \emph{i.e.} it has an infinite entropy cost) so that a
change of energy reference does not affect the value of $K$.
Equation~\ref{eq:omega-E} leads us to recover the parabolic
approximation for the microcanonical configurational entropy
\begin{equation}
 S_c(E)=S_{\infty}-\frac{(E-E_{\infty})^{2}}{2K}
\label{eq:micro}
\end{equation}
which takes the same form as that already introduced in
Eq.~\ref{eq:conf-sim} to phenomenologically describe simulated
liquids. Finally, using Eq.~\ref{eq:omega} we can also predict how the
average structural composition of the liquid evolves when the energy is
lowered:
\begin{equation}
  \avv{\vec{c}}(E) = \frac{1}{\Omega(E)} \int \ \vec{c}\ \  \delta( E - \vec{c} \cdot \vec{\epsilon} ) \  \Omega(\vec{c}) \ \mathrm{d}\vec{c}
  \label{eq:c-E-interm}
\end{equation}
which yields:
\begin{equation}
  \avv{\vec{c}}(E) = \vec{c}_\infty + \frac{E-E_\infty}{K}\hat{C}\cdot \vec{\epsilon}
  \label{eq:c-E}
\end{equation}
This shows that the evolution of the concentration of
structure $i$ depends not only on its energy $\epsilon_i$, but also on
the energy of other structures $j$ through the geometric coupling due
to their overlap:
\begin{equation}
  \frac{\partial \avv{c_i}}{\partial E}  =  \frac{1}{K} \sum_j C_{ij} \  \epsilon_j
  \label{eq:dcdE}
\end{equation}
In particular, this shows that the accumulation of a given local
structure upon cooling is not equivalent to it being stable: some
possibly unfavoured structures will tend to accumulate due to their
geometrical affinity to other stable local structures.

Our framework thus provides an explicit connection between the local
structural properties of a liquid and its thermodynamic properties.
The main outcome of our reasoning is the analytic prediction of
Eq.~\ref{eq:K} for the quadratic coefficient $K$, which is the key
quantity controlling the entropic cost of lowering the potential
energy of the liquid. It combines the data of the local energy
landscape -- the energy vector $\vec{\epsilon}$, which can be
straightforwardly computed from the set of local structures -- with
the geometrical properties of structures -- the covariance matrix
$\hat{C}$ encoding symmetry and overlap properties of local structures
-- to give a simple number that predicts the high-temperature
properties of the system. Using these quantities, we are also able to
predict, through Eq.~\ref{eq:dcdE}, how the liquid's structure will
change on cooling.

Note that the truncation of the expansion of the entropy to second
order means that we only account for the correlations between pairs of
local structures, and neglect correlations between three or more local
structures. As an example of such higher order correlations neglected
by this approximation, Cheng et al~\cite{ma-icos} have reported the
existence of extended arrangements of local polyhedra such as rings of
icosahedra. It would be in principle feasible to extend our framework
to higher orders; however it would increase dramatically its
complexity.

Central to our program is the calculation of the infinite-temperature
structural covariance matrix $\hat{C}$, which encodes all pertinent
information about geometrical interactions of pairs of structures.  It
could already be empirically obtained through Eq.~\ref{eq:C} provided
we have knowledge on the high-temperature inherent states
statistics. However, we suggest that it could also be derived from
first principles, using only the geometrical properties -- symmetry
and pair overlap -- of the local structures. This program, which we
have already carried on in the case of lattice
liquids~\cite{ronceray6}, would allow a quantitative approximation of
the liquid configurational entropy in terms of purely local and
geometrical properties of the liquid.

\section{ Applying the Structural Covariance Analysis to Lattice Spin
  Models} \label{sec:fls}

The application of the program outlined in the previous Section to a
simulated liquid with continuous degrees of freedom has not yet been
carried out. In its place, we present in this Section an application
of the theory to a lattice model of liquids, the Favoured Local
Structure (FLS) model. Our goal is to demonstrate the applicability of
the structural covariance approach to computing the configurational
entropy, and investigate the range of temperatures over which it
provides physically relevant results. The idea of this discrete model
is that it provides a finite set of local structures, over which we
can choose explicitly an arbitrary local energy landscape. We start
this Section by presenting a generalized version of the FLS model, and
showing how to implement our structural covariance approach within
this model.  We finally extend this model to the case of multiple FLS,
and show how our high-temperature covariance analysis provides
precious information about the glass-forming ability of the liquid.


\subsection{The Favoured Local Structures Model}
\label{sec:FLS}

\begin{figure}[tb]
  \centering
  \includegraphics[width=\linewidth]{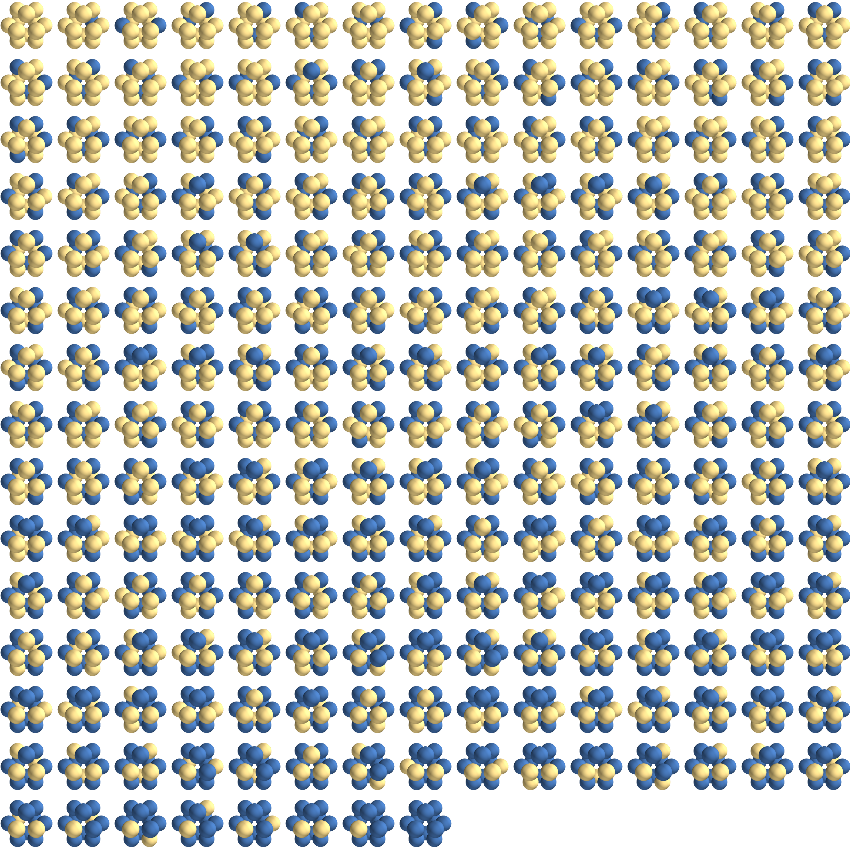}
  \caption{The $218$ local structures at the nearest neighbour level
    in a binary spin model on the face-centered cubic lattice, ordered
    by increasing number of up (blue) spins. Note that here we do not
    consider the value of the central spin in our description; doing
    so would double the number of structures.}
  \label{fig:all_FLSs}
\end{figure}

The Favoured Local Structure (FLS) model consists in two-state spins
on a face-centered cubic (FCC) lattice, in which we explicitly define
the Hamiltonian of the system in terms of a simple local energy
landscape.  Each site has $12$ neighbours and on each site we have a
spin that is either up or down. Among the $2^{12} = 4096$ possible
combinations for the spin values of these $12$ neighbours, we identify
$218$ geometrically distinct local structures, which are not related
by to one another by a rotation, as depicted in
Figure~\ref{fig:all_FLSs}. The principle of our model is to attribute
an energy $\epsilon_i$ to each local structures $i$ of this
set. Therefore, denoting $c_i$ the proportion of sites of a
macroscopic system in this local structure, the average energy per
site reads:
\begin{equation}
  \label{eq:E_FLS}
  E = \sum_{i=1}^{218} c_i \epsilon_i = \vec{c}\cdot \vec{\epsilon}
\end{equation}
such that our model fulfills precisely the assumptions of
Section~\ref{sec:theory}, where the inherent structures of the liquid
are now replaced by spin configurations of the model.

An instance of the FLS model thus corresponds to the choice of the
energy vector $\vec{\epsilon}$ in the huge, 218-dimensional parameter
space. Note that we can select these energies arbitrarily, without
requiring that they derive from a pair potential. This
convenient property allows us to study directly the influence of the
geometry of local structures, without being concerned with finding
fine-tuned interaction potentials that would favour these structures.
Once this \emph{local energy landscape} $\vec{\epsilon}$ is set,
Eq.~\ref{eq:E_FLS} fully defines the Hamiltonian of the system. We
typically study this model numerically, employing the Metropolis
algorithm in canonical ensemble simulations of a finite system of size
$30^3$ lattice sites, with periodic boundary conditions. Note that we
do not enforce conservation of the total spin.

The aim of this model is to gain insight on the influence of locally
favoured geometries on the thermodynamic properties of the system. A
full study of this gigantic parameter space would be both daunting and
pointless; instead, we focus on reduced versions with only a few
non-zero $\epsilon_i$'s. We have recently presented an extensive
study~\cite{ronceray5} of the case of a single favoured local
structure $i$, \emph{i.e.} with $\epsilon_j = -\delta_{ij}$ where
$\delta_{ij}$ is the discrete delta function. In that case, each site
in the FLS has an energy $-1$, while all others have an energy $0$,
and the differences between systems can be completely attributed to
the geometrical properties of the FLS. In particular, we showed that
low-symmetry local structures tend to stabilize better the liquid than
high-symmetry ones. Indeed, their entropy cost is lower due the
remaining uncertainty on the orientation of low-symmetry local
structures.

The other key component to the liquid's stability with respect to
crystallization is the ground state energy $E_o$, corresponding in
this model to a dense packing of FLS's on the lattice. In our model,
this energy ranges from $E_o = -1$ (for those FLS's which can tile the
whole space, such as the all-up local structure) to $E_o=-0.25$,
corresponding to a highly \emph{frustrated} local order where only one
site out of four, at most, can be in the FLS, i.e. geometrical constraints
prevent the local energy minimum from extending to the whole space,
and effected termed \emph{geometrical frustration}~\cite{frust}.
While the connection between FLS symmetry and ground state energy
turns out to be complex, we have observed that \emph{chiral} FLS's --
\emph{i.e.} local structures lacking a plane reflection symmetry --
tend to be much more frustrated than achiral ones. As we will show
here, we can use this property to generate glass-forming systems by
combining several highly frustrated FLS's.

\subsection{Structural covariance analysis of the FLS model}
\label{sec:FLS-cov}

We now apply the theoretical analysis presented in
Section~\ref{sec:theory} to this model, for any choice of the local
energy landscape $\vec{\epsilon}$. Again, we first consider the
entropy in the structure space $S(\vec{c})$, which is a generic
property of our model, independent of $\vec{\epsilon}$.

The infinite-temperature structural composition of the liquid is
simply given by the symmetry properties of local structures: each
local structure $i$ can fit at any lattice site in $g_i = s_i/\omega$
distinct orientations, where $\omega = 24$ is the size of the rotation
group of the FCC lattice and $s_i$ is the cardinal of the rotation
subgroup that leaves the structure unchanged. At infinite temperature,
the spins are completely independent, and all $2^{12}$ local spin
environments are equiprobable; hence $c_{i,\infty} = g_i/2^{12}$.

\begin{figure}[tb]
  \centering
  \includegraphics[width=\linewidth]{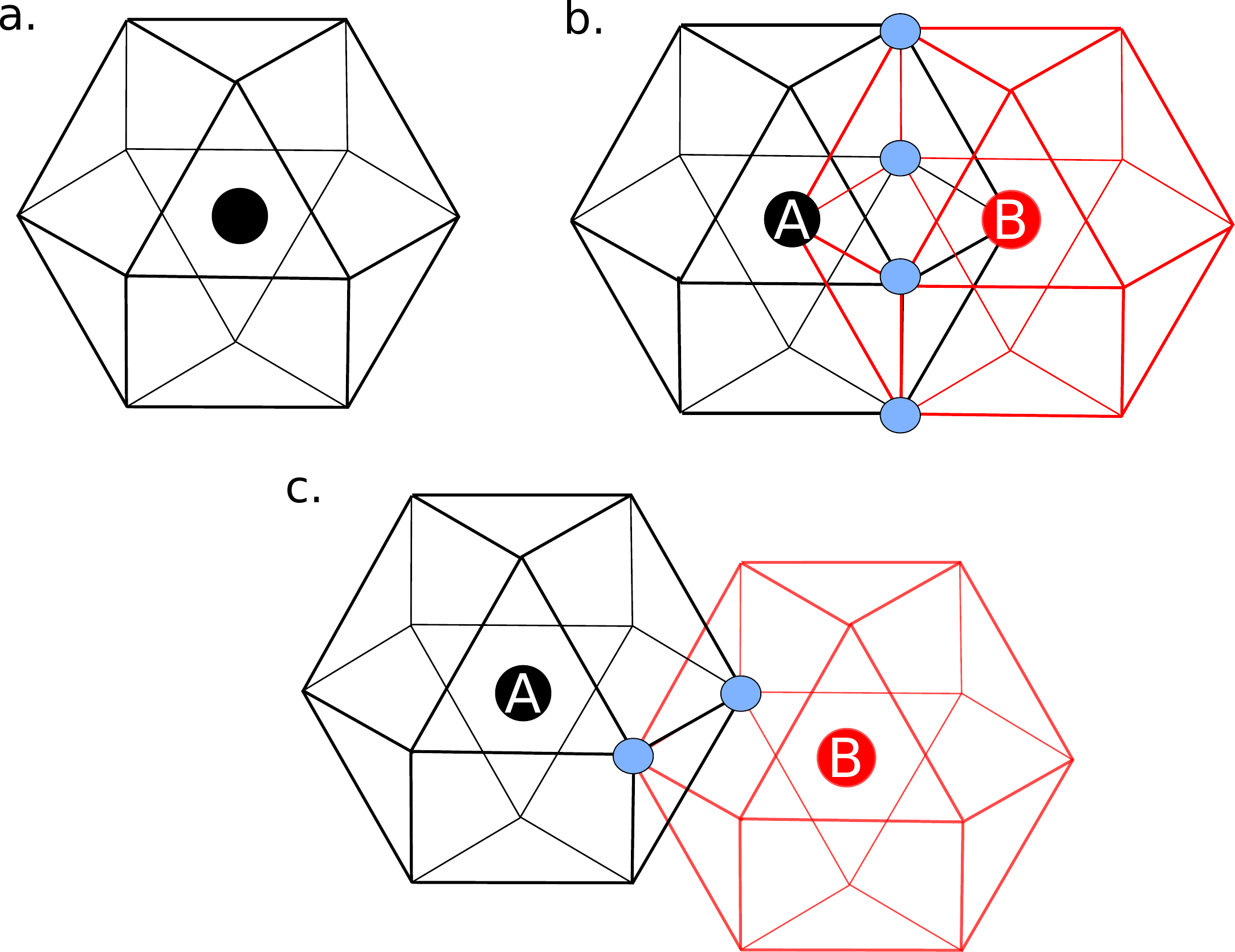}
  \caption{(a). The local structure at a lattice site (black circle) is
    determined by the spin values of the sites of its coordination
    shell which has the geometry of a cuboctahedron. (b) and (c). The
    local environments of two neighbouring sites A and B overlap at 4 points (blue circles in (b)), 2 point (blue circles in (c)) or 1 point. These overlaps impose mutual constraints on the local structures that sites A and B can adopt. }
  \label{fig:overlap}
\end{figure}

Expanding the entropy $S(\vec{c})$ around the infinite-temperature
composition $\vec{c}_\infty$ as in Eq.~\ref{eq:entropy} defines the
infinite-temperature covariance matrix $\hat{C}$ of structural
compositions in this model (Eq.~\ref{eq:C}). This matrix is a purely
geometrical property of the set of structures presented in
Fig.~\ref{fig:all_FLSs}.  As discussed in the previous Section, this
matrix provides a measure of the degree to which each possible pair of
local structures act to constrain each other. The physical origin of
this constraint is local, arising from the geometrical restrictions
involved in the overlap of two coordination shells. In our model, this
overlap consists in lattice sites being shared between adjacent local
structures, as shown in Fig.~\ref{fig:overlap}. A complete enumeration
of these local constraints allows us to compute exactly the covariance
matrix $\hat{C}$, as detailed further in Ref.~\cite{ronceray6}. The
principle of this calculation is to consider, for each pair of
neighbouring lattice sites such that the local structures overlap, the
change in infinite-temperature probability of observing structure $i$
at one site knowing that structure $j$ is present at the other. If
structures $i$ and $j$ are geometrically incompatible, this will be
impossible and the probability will be zero, while if they are
compatible the shared sites result in an enhanced probability.

\begin{figure}[tb]
  \centering
  \includegraphics[width=\linewidth]{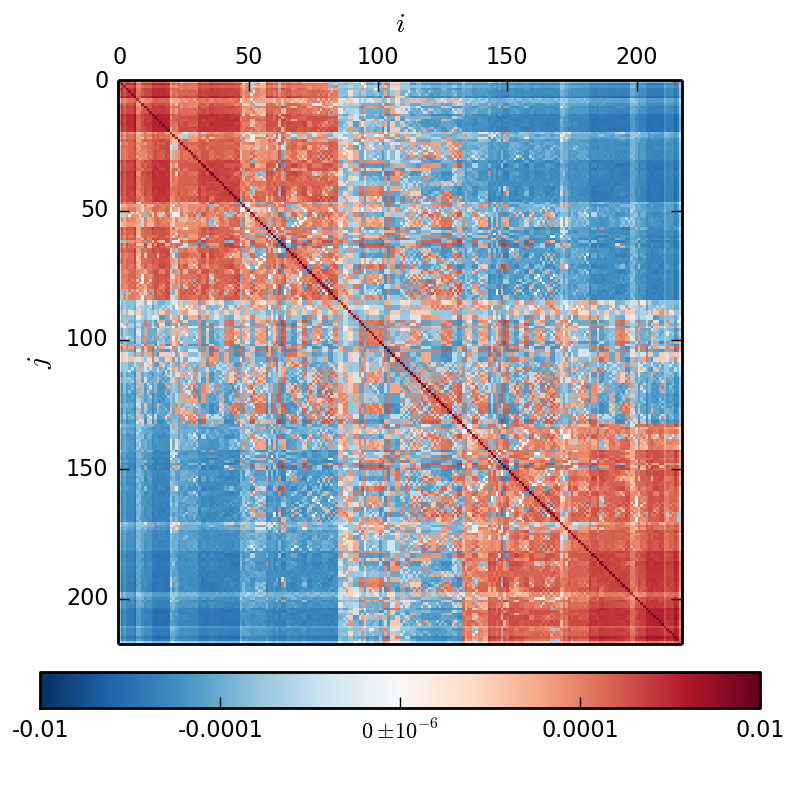}
  \caption{A colour coded depiction of the values of the elements of
    the covariant matrix $\hat{C}$. Negative and positive elements are
    in shades of blue and red, respectively. These shades are
    log-scaled with a cut region around zero, as indicated by the
    colorbar. The local structures, listed by a numerical index
    running from 1 to 218, are sorted by their number of up spins as
    in Figure~\ref{fig:all_FLSs} (hence the block structure of this
    matrix, as local structures with similar spin composition will
    tend to attract each other). Within each block, they are sorted
    with increasing geometrical multiplicity $g_i$.}
  \label{fig:Cij}
\end{figure}

In Figure~\ref{fig:Cij}, we present a graphical representation of the
high temperature covariance matrix $\hat{C}$. We emphasize the fact
that this matrix does not depend on a particular choice of the
energies of the different structures (\emph{i.e.} the choice of
$\vec{\epsilon}$) as it is entirely determined by the geometrical
constraints associated with overlapping coordination shells.

Combining the covariance matrix with the energy vector, we find again
the quadratic approximation formula for the microcanonical entropy,
\begin{equation}
  S(E) = S_\infty - \frac{1}{2K}(E-E_\infty)^2
  \label{eq:S-repeated}
\end{equation}
where the thermodynamic coefficient $K$ is given by Eq.~\ref{eq:K}.
In the case of the single FLS model, with $\epsilon_j = -\delta_{ij}$,
this coefficient becomes
\begin{equation}
  K=\vec{\epsilon}\cdot\hat{C}\cdot\vec{\epsilon} = C_{ii}
  \label{eq:K-FLS}
\end{equation}
Thus only self-overlap of the FLS is important in this quadratic
approximation, while all other local structures are spectator to the
accumulation of order in the liquid. In spite of its simplicity, the
version of the FLS model characterized by Eq.~\ref{eq:K-FLS} has
proven useful in understanding some of the factors that influence the
temperature dependence of the liquid entropy~\cite{ronceray2}.

Because in this model the total entropy equals the configurational
entropy, Eq.~\ref{eq:S-repeated} can be used to map our results to
finite temperature properties through the formula $\partial S
/ \partial E = 1/T$. Our framework thus provides predictions for the
temperature dependence of the entropy and energy of the system:
\begin{eqnarray}
  E(T) &= E_{\infty}-K/T \label{eq:EK}\\
  S(T) &=  S_{\infty}-\frac{K}{2T^{2}} \label{eq:SK}
\end{eqnarray}
Similarly, adapting Eq.~\ref{eq:c-E} yields
\begin{equation}
  \avv{\vec{c}}(T) = \vec{c}_\infty  -\frac{1}{T}\hat{C}\cdot \vec{\epsilon}
  \label{eq:c-T}
\end{equation}
Equations~\ref{eq:EK}-\ref{eq:c-T} become exact in the
high-temperature limit, as they correspond to an exact expansion up to
$O(T^{-2})$. They thus provide a quantitative approximation for the
liquid's thermodynamic properties and structural composition at finite
temperature.

\subsection{Multiple FLS and Glass-Forming Ability}
\label{sec:multiFLS}

We now present a new variant of this model, which has the specificity
to present glass-forming systems, and use our covariance formalism to
study it.

\begin{figure}[tb]
  \centering
  \includegraphics[width=0.7\linewidth]{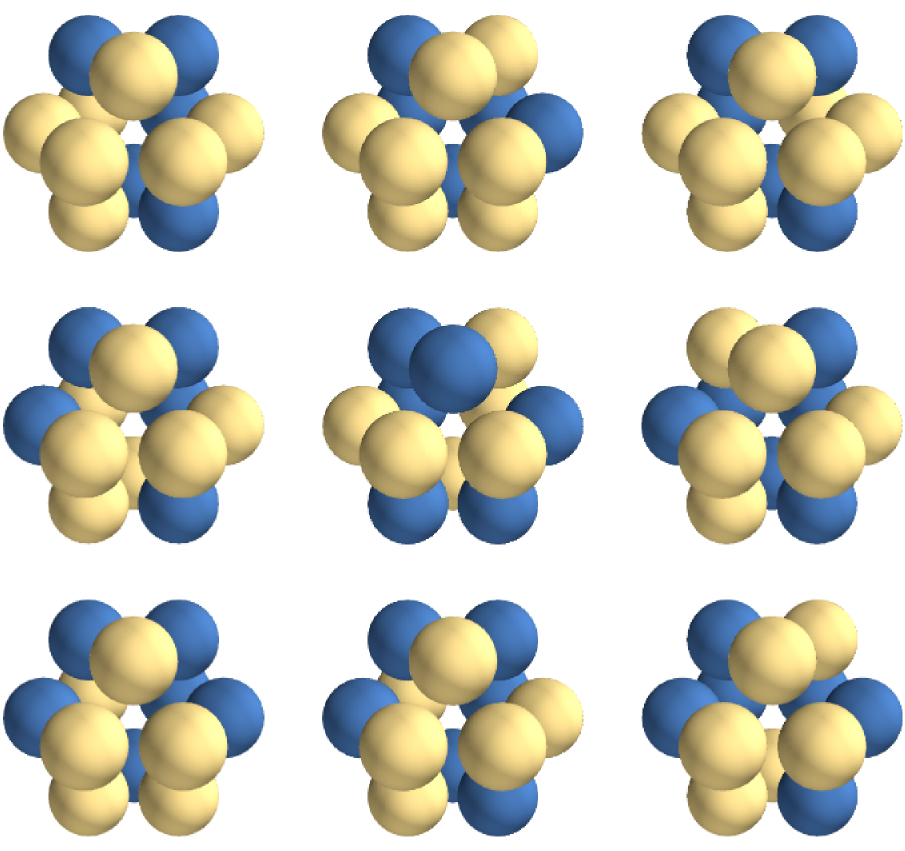}
  \caption{Depiction of the nine FLS's we consider for our model with
    two favoured structures. These are denoted respectively (from left
    to right and top to bottom) "43c", "47c", "51c", "53c", "62c",
    "63c", "69c", "70c" and "71c" in our comprehensive study of
    single-FLS cases~\cite{ronceray5}. }
  \label{fig:LSs}
\end{figure}

Let us consider the more general -- and perhaps more realistic --
scenario in which a liquid can be characterized by multiple FLS's. We
model this using our lattice model by attributing an energy
$\epsilon_i = -1$ to a number of local structures (while, again,
setting the remainder to zero energy)~\cite{ronceray6}.  The number of
options for such liquids in the FLS model is huge -- $47306$
possibilities for only two FLS's in the 3D FCC model -- so we shall
restrict ourselves here to constructing bi-FLS liquids from a specific
subset of local structures, as depicted in Fig.~\ref{fig:LSs}. These
structures have been chosen because, in the single FLS system, they
have a high ground state energy above $-0.3$, meaning that at most
$30\%$ of the lattice sites can host a FLS in a given configuration of
the system. This high level of frustration results in a low
equilibrium freezing temperature, as the crystalline groundstate is
not very stable, and therefore leads to more interesting liquid
physics. Besides, the structures of Fig.~\ref{fig:LSs} are chosen to
have either $5$ or $6$ up spins, such that no ferromagnetic alignment
effect takes place: only the geometrical features of the selected
structures are important.

We consider here bi-FLS systems in which two of these local structures
are favoured with an energy $-1$, while all other structures have zero
energy.  Note that each of our nine selected local structures are
chiral, \emph{i.e.}  they lack a reflection symmetry.  We also include
spin-inverted and mirror-symmetric cases for these structures, which
have the same properties as the original FLSs, but different overlap
properties with each other.  This results in a pool of $126$ possible
pairings of such chiral, high ground state energy systems, which we
have investigated comprehensively using Monte-Carlo simulations.

\subsubsection{Scenarios for the low-temperature fate of the liquid}
\label{sec:scenarios}

\begin{figure*}[tbp]
  \centering
  \includegraphics[width=0.8\linewidth]{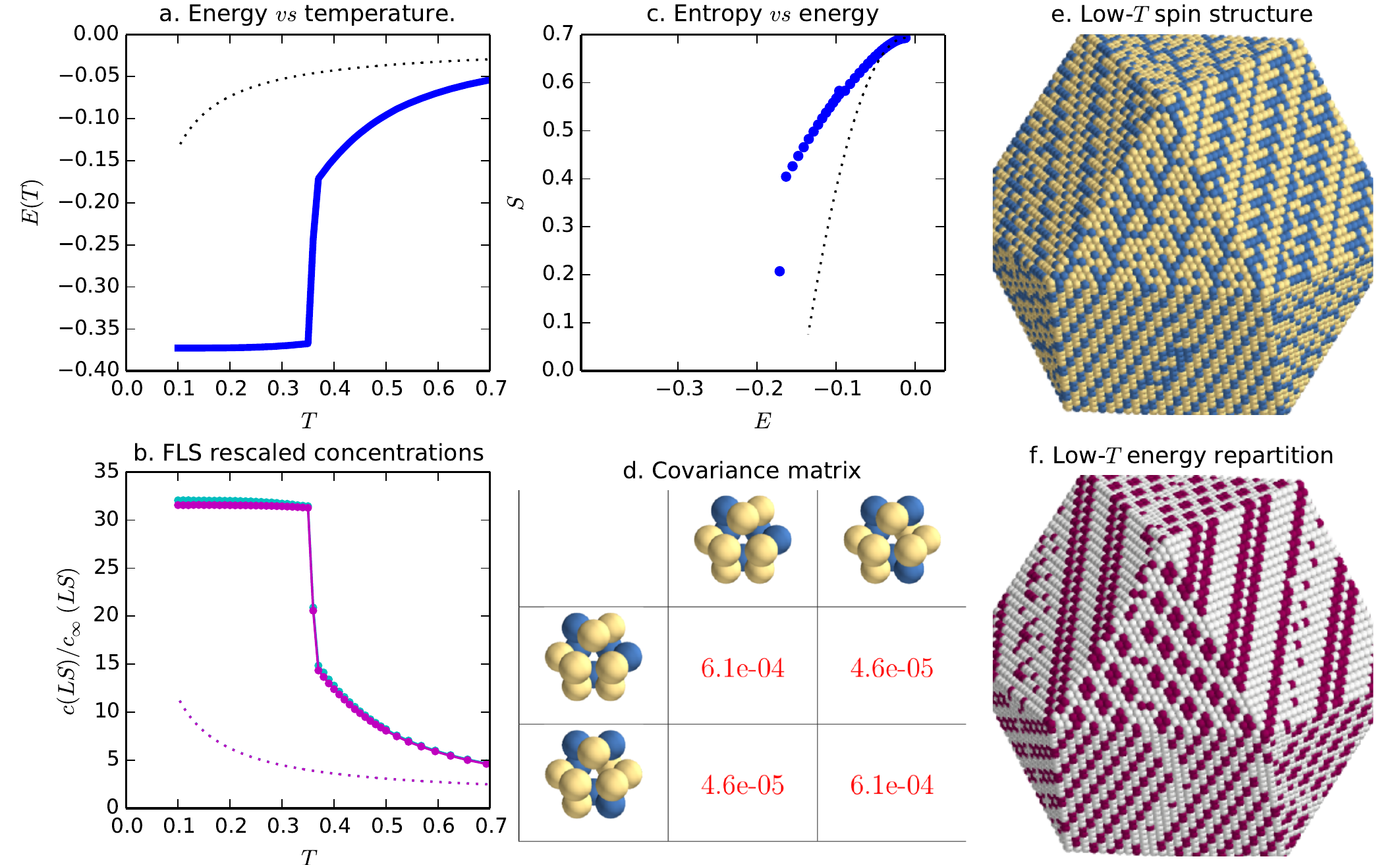}
  \caption{An example of a bi-FLS model that crystallizes. Structures
    $47c$ and $51c$ are both favoured. \textbf{a.} The average energy
    per site exhibits a sharp first-order phase transition on
    cooling. Dotted line: the high-$T$ expansion prediction
    (Eq.~\ref{eq:EK}). \textbf{b.} The concentration of the two FLS's
    (filled circles of purple and blue) as a function of
    temperature. The concentrations predicted by the high $T$
    approximation (\emph{i.e.} Eq.~\ref{eq:c-T}) are also included
    (dotted line) for comparison. \textbf{c.} The liquid entropy per
    site as a function of the energy. Also plotted, for comparison, is
    the high-$T$ quadratic form (Eq.~\ref{eq:S-repeated}, dotted line)
    which becomes exact near $E_\infty$. \textbf{d.}  The two favoured
    local structures and their associate high-$T$ covariance. The
    positive off-diagonal terms indicate geometric affinity between
    these structures. \textbf{e.}  Slices of a low-temperature
    configuration reveal a crystalline structure. \textbf{f.} The loci
    of FLS (red spheres) are regularly organized. Parameters: system
    size $30\times 30 \times 30$ with periodic boundary conditions;
    cooling rate $10^6$ Monte-Carlo steps/site/temperature unit.}
  \label{fig:scenario-crystal}
\end{figure*}

\begin{figure*}[btp]
  \centering
  \includegraphics[width=0.8\linewidth]{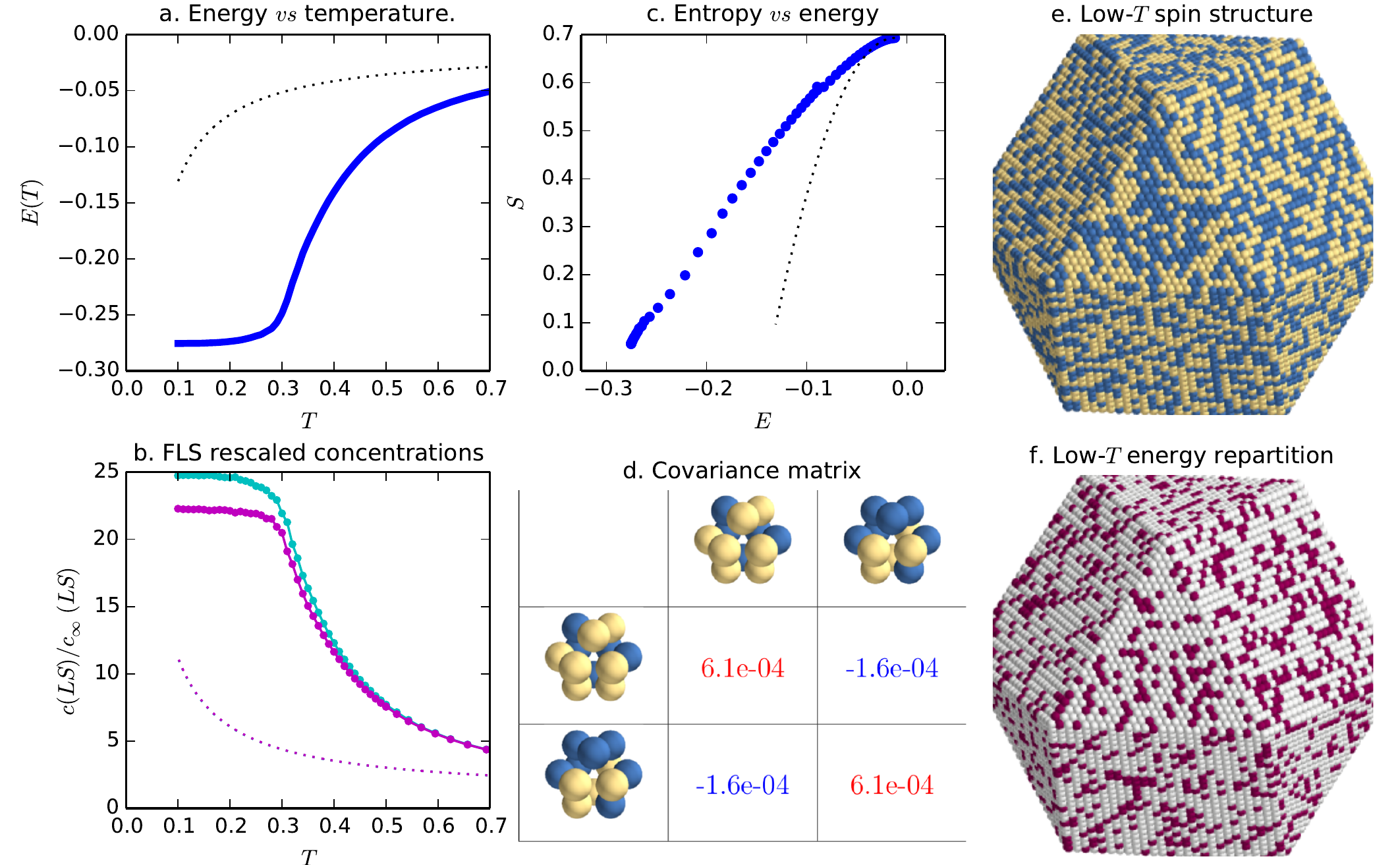}
  \caption{An example of a glass-forming bi-FLS system: structures
    $47c$ and the spin-inverted, mirror-symmetric version of $51c$ are
    both favoured. \textbf{a.}  The average energy per site exhibits a
    continuous decrease on cooling, down to the point of dynamical
    arrest. Dotted line: the high-$T$ expansion prediction
    (Eq.~\ref{eq:EK}).  \textbf{b.} The  concentrations of the
    two FLS's (filled circles of purple and blue) as a function of
    temperature. \textbf{c.} The liquid entropy per site as a function
    of the energy. Also plotted, for comparison, is the high-$T$
    quadratic form (Eq.~\ref{eq:S-repeated}, dotted line). \textbf{d.} The
    two favoured local structures and their associate high-$T$
    covariance. The negative off-diagonal terms indicate geometric
    repulsion between these structures. \textbf{e.} Slices of a
    low-temperature configuration reveal an absence of
    crystallinity. \textbf{f.} The loci of FLS (red spheres) are
    apparently random. Parameters: system size $30\times 30 \times 30$
    with periodic boundary conditions; cooling rate $10^6$ Monte-Carlo
    steps/site/temperature unit.}
  \label{fig:scenario-glass}
\end{figure*}

We can identify two distinct scenarios in the behaviour of these
liquids on cooling: they either crystallize
(Figure~\ref{fig:scenario-crystal}), or continuously arrest into a
disordered, glassy state (Figure~\ref{fig:scenario-glass}).

\textbf{Crystallization} (79 systems). In the majority of studied
bi-FLS systems, the observed scenario is the crystallization of the
system on cooling. An example of this behaviour in bi-FLS systems is
illustrated in Fig.~\ref{fig:scenario-crystal}. On cooling, the energy
decreases (Figure~\ref{fig:scenario-crystal}\textbf{a}) as FLS's
accumulate (Figure~\ref{fig:scenario-crystal}\textbf{b}). Freezing is
marked by an abrupt first order transition in these variables.  This
results in the end of the liquid branch in the $S(E)$ microcanonical
plot
(Figure~\ref{fig:scenario-crystal}\textbf{c}). Figure~\ref{fig:scenario-crystal}\textbf{d}
shows the covariance sub-matrix characterizing geometrical
interactions between the FLS's in the system.  At temperatures below
the freezing point, the system is in a low energy state with periodic
arrangement of both spins
(Figure~\ref{fig:scenario-crystal}\textbf{e}) and the FLS sites
(Figure~\ref{fig:scenario-crystal}\textbf{f}).  The results of the
high temperature approximation for $E(T)$, the FLS concentrations and
$S(E)$ are also plotted (dotted lines). We find that the approximation
increasingly underestimates the concentration of FLS and the liquid
entropy as the temperature and energy, respectively, decrease. This
shortcoming of the high $T$ approximation is quite generic for the FLS
model and suggests the non-trivial result that the inclusion of higher
order correlations between local structures serves to \emph{decrease}
the entropy cost of the FLS.

\textbf{Glass Transition} (47 systems). In more than a third of cases,
crystallization is avoided, and the slow cooling of the system down to
zero temperature leaves us with a dynamically arrested disordered
state.  We refer to this outcome as a glass transition, and illustrate
it with an example bi-FLS system in
Figure~\ref{fig:scenario-glass}. The system continuously accumulates
the two FLS's with decreasing $T$. We note that the degeneracy of
these two local structures breaks down as the arrested state is
approached (see Figure~\ref{fig:scenario-glass}\textbf{b}).  The
temperature dependence of the energy
(Figure~\ref{fig:scenario-glass}\textbf{a}) and the variation of the
entropy with energy (Figure~\ref{fig:scenario-glass}\textbf{c})
characterise a continuous transformation into the disordered structure.  A
small heat capacity peak is associated with this transition which, in
preliminary calculations, does not exhibit any systematic increase
with system size. The absence of a dependence on system size suggests
that there are no long range correlations associated with the observed
glass transition. Note that among the 126 systems we studied, a few are somewhat
ambiguous, as they undergo continuous accumulation of FLS's yet
exhibit structured patterns that could be crystalline.

\subsubsection{Structural covariance and low-$T$ fate of the liquid}
\label{sec:correl_Cij}

In these bi-FLS systems with favoured local structures $i$ and $j$,
the high-temperature expansion gives a simple form for the
thermodynamic constant $K$ characterizing a system:
\begin{equation}
\label{eq:K-biFLS}
K = C_{ii} + C_{jj} + 2 C_{ij}
\end{equation}
The diagonal terms $C_{ii}$ and $C_{jj}$ characterize self-overlap of
the local structures ~\cite{ronceray5}. Structures that overlap well
with themselves have larger $C_{ii}$ and accumulating them is less
costly in entropy. Positive by construction, the diagonal terms are
also dominated by the symmetry properties of the FLS's so that
low-symmetry FLS's cost less entropy, and result in larger values of
$K$.

\begin{figure}[ht]
  \centering
  \includegraphics[width=0.8\linewidth]{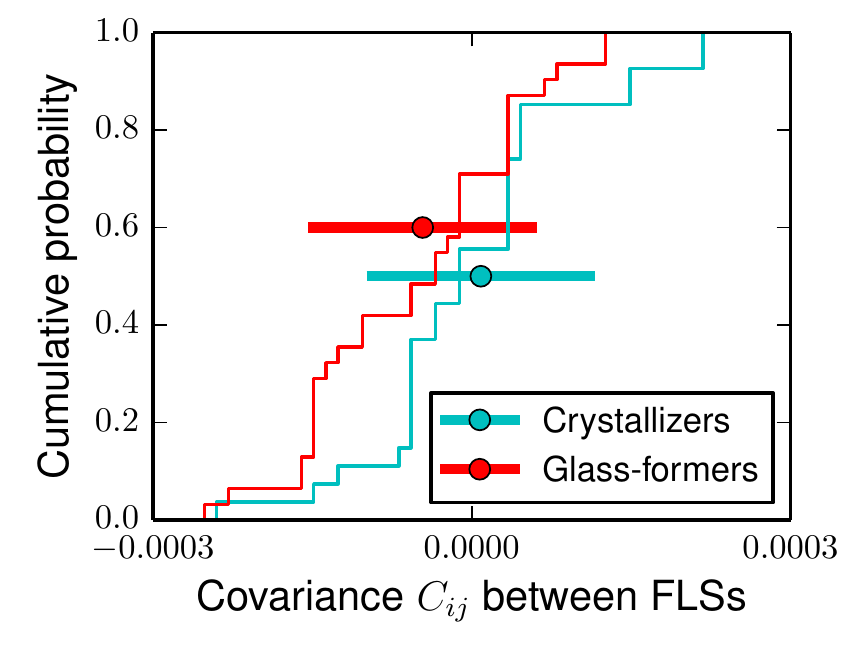}
  \caption{ The cumulative probability of the structural covariances,
    for the $47$ glass-forming and the $79$ crystallizing systems. The
    filled circles with error bars indicate the mean and standard
    deviation, respectively, within each population. }
  \label{fig:Cij-cumul}
\end{figure}

The off-diagonal elements of the covariance matrix capture the
entropic interactions between different FLS. For the examples
presented in Figures~\ref{fig:scenario-crystal}
and~\ref{fig:scenario-glass}, we present the physically relevant
portion of the covariance matrices
(Figures~\ref{fig:scenario-crystal}\textbf{d} and
\ref{fig:scenario-glass}\textbf{d}). In the crystallizing example, we
have $C_{ij} = 4.6\times 10^{-5}$, indicating that the two FLS's have
positive overlap, the presence of one at a given site tends to
increase the probability to find the other in its surroundings. In
contrast, in the glass-forming example, we have $C_{ij} = -1.6\times
10^{-4}$, indicating an effective repulsion between local structure,
the presence of one depletes the concentration of the other in the
neighbourhood. This trend, $C_{ij}>0$ for crystallizers and $C_{ij}<0$
for glass formers, does not apply to every bi-FLS choice but, over our
entire sample of 126 systems, is statistically significant. The
average values of the off-diagonal covariance $C_{ij}$ for the
crystallizers and the glass formers are, respectively,
\begin{eqnarray*}
\langle C_{ij} \rangle_\mathrm{crystal} = & +0.7\times 10^{-5} \pm 7\times 10^{-5} \\
\langle C_{ij} \rangle_\mathrm{glass} = & -3.5\times 10^{-5} \pm 1\times 10^{-4}
\end{eqnarray*}
Glass-forming ability therefore correlates with poor overlap between
structures and hence geometric repulsion, while crystallizers are in
general characterized by a geometric attraction between FLS's.  This
effect is statistically significant, a two-tailed Student $t$-test on
the value of $C_{ij}$ for the populations of glass-formers and
crystallizers indicates a $p$-value of $p=0.007$.  This correlation
has however a rather low signal-to-noise ratio, as shown in
Figure~\ref{fig:Cij}, where the distributions of the structural
covariances overlap between the two populations. Although the
correlation is weak, the fact that a discernable correlation exists is
significant, as it supports our proposal that the high $T$ structural
correlations contains information about the fate of the low $T$
liquid.

\subsubsection{The role of crystal groundstate energies}
\label{sec:Eo}

\begin{figure}[ht]
  \centering
  \includegraphics[width=0.8\linewidth]{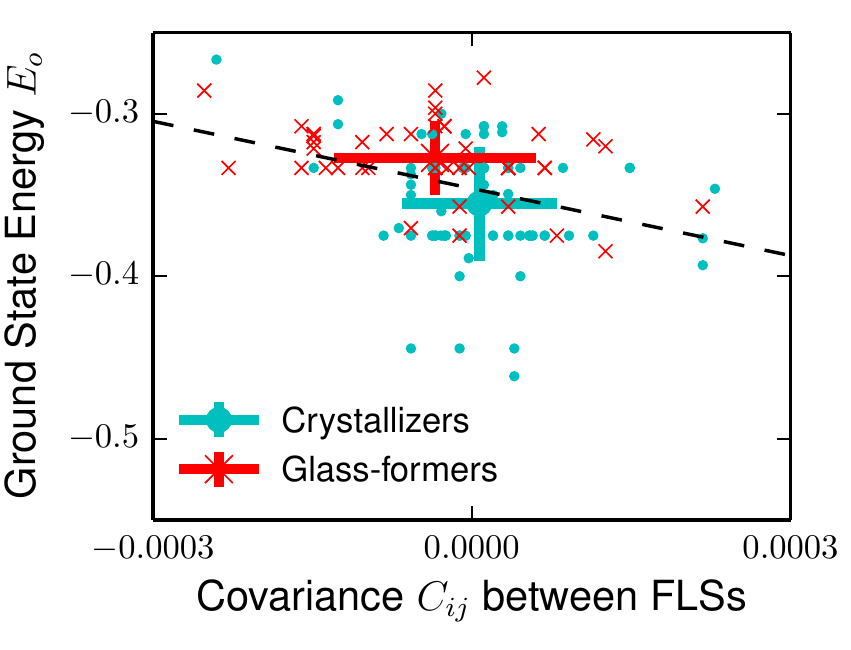}
  \caption{ Scatter plot of the values of the ground state energy
    $E_o$ as a function of structural covariance $C_{ij}$ for all 126
    bi-FLS systems. Dots: crystallizing systems, crosses:
    glass-forming systems. The large symbols with error bars indicate
    the mean and standard deviations within each population. The black
    dashed line is a linear regression.  }
  \label{fig:EoCij}
\end{figure}

While we have established that the high $T$ covariance between FLS's
contains information relevant to the low-temperature fate of a liquid,
the weakness of the correlation indicates that higher order entropic
interactions between FLS, missing from the high $T$ expression, are
important. One accessible source of information on the organization of
FLS at low $T$ is the crystal groundstate. This state represents the
densest packing of the selected FLS on the FCC lattice. Establishing
the crystal groundstates for 126 different Hamiltonians via structural
searches is a substantial challenge. We have applied the lattice
search algorithm described in ref.~\cite{ronceray5}, an enumerative
method encompassing all possible crystalline structures with crystal
cell size $\mathcal{Z}\leq 64$ and complemented by large-scale
Monte-Carlo simulations. Since the same overlap constraints in the
bi-FLS models influence the groundstate energy and the high $T$
covariance $C_{ij}$ we would expect some correlation between the two
quantities.  We test this conjecture in Figure~\ref{fig:EoCij} where
we plot the scatter of the pairs $(C_{ij},E_o)$ for all $126$ systems
studied. We find a negative correlation between $E_o$ and $C_{ij}$,
with Pearson correlation coefficient $r = -0.34$ and significance
$p=10^{-4}$. Again, this correlation is rather weak, but significant.
High-temperature covariances correspond to simple combinatorics that
do not encompass the whole complexity of dense packing problems but
they serve as useful indicators of whether FLS's will pack well or
not.

\begin{figure}[ht]
  \centering
  \includegraphics[width=0.8\linewidth]{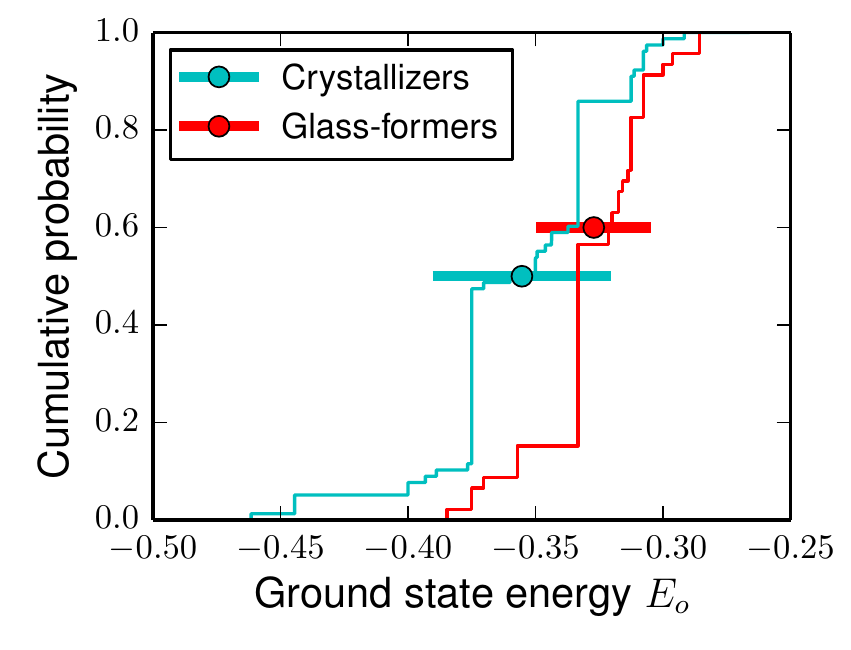}
  \caption{ The cumulative probability of the ground state energies
    $E_o$. The filled circles with error bars indicate
    the mean and standard deviation, respectively, within each population. }
  \label{fig:Eo}
\end{figure}

$E_o$ also correlates with glass-forming ability.  High ground state
energies result in stabilization of the liquid with respect to
crystallization and, therefore, in low the crystallization
temperatures. This in turn increases the likelihood that the liquid
will dynamically arrest before freezing. Averaging the ground
state energies with crystallizing and glass-forming systems, we find:
\begin{eqnarray*}
  \langle E_o \rangle_\mathrm{crystal} = &   -0.355 \pm 0.04 \\
  \langle E_o \rangle_\mathrm{glass} = &  -0.327   \pm 0.02
\end{eqnarray*}
\emph{i.e.}, the former have significantly lower ground state energies
than the latter. A Student $t$-test indicates a strong significance
for this trend, with $p=10^{-6}$.  We thus conclude that a possible
mechanism explaining why geometric repulsion between favoured local
structures correlates with glass-forming ability is through the high
ground state energy it incurs.

\subsubsection{What determines whether a liquid will crystallize or form a glass?}
\label{sec:Eo}

The question of what factors influence the fate of the liquid is of
central importance both for the development of new bulk metallic
glasses~\cite{bmg} and in terms of understanding the fundamental
question of the low temperature end of the liquid state.  The kinetics
of crystallization is obviously of central importance. The
relationship between crystal nucleation and growth and the structural
statistics of the liquid is a substantial problem that we leave for
future work. Here we restrict ourselves to consider a key parameter of the
crystallization kinetics: the equilibrium freezing temperature. The
lower this temperature, the slower the relaxation of the system at the
freezing point, and hence the higher the likelihood to miss
crystallization and form a glass.

\begin{figure}[ht]
  \centering
  \includegraphics[width=0.8\linewidth]{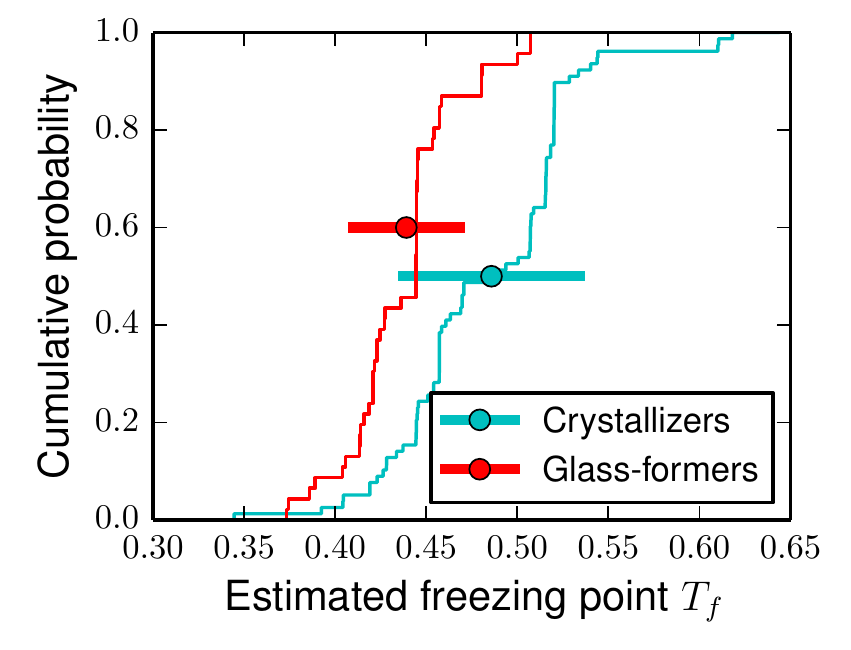}
  \caption{ The cumulative probability of the ground state energies
    $E_o$. The filled circles with error bars indicate
    the mean and standard deviation, respectively, within each population.  }
  \label{fig:Tf}
\end{figure}

 Previously~\cite{ronceray6}, we derived an expression for the freezing
temperature $T_f$ based on the high $T$ expression for the entropy,
\begin{equation}
T_{f} =  \left({\frac{E_{\infty}-E_o}{2S_{\infty}}}\right) \left(1+\sqrt{1-\frac{2KS_{\infty}}{(E_{\infty}-E_o)^{2}}}\right)
\label{eq:tf}
\end{equation}
Here $E_{\infty}$ and $S_{\infty}$ are the high $T$ limit of the energy and entropy per site, respectively. In Figure~\ref{fig:Tf} we
plot the cumulative distributions of this estimated value for $T_f$ for the glass
forming and crystallizing bi-FLS models. While the two distributions
still overlap, they appear to be better separated by
$T_f$. Quantitatively, we find:
\begin{eqnarray*}
  \langle T_f \rangle_\mathrm{crystal} = &   0.49 \pm 0.05 \\
  \langle T_f \rangle_\mathrm{glass} = &  0.44   \pm 0.03
\end{eqnarray*}
and the $p$-value is now $p=10^{-7}$, lower than for both $C_{ij}$ and
$E_o$.  By combining information of both the high $T$ structural
correlations and the $T=0$ the groundstate energy $E_o$ into an
estimated freezing temperature, we thus obtain a better predictor of
the glass-forming ability of a system than by the covariance $C_{ij}$
alone. This result is consistent with the experimental
observation~\cite{ping} that a low freezing point is one of the best
indicators of glass forming ability.

\section{Discussion}

In this article, we have proposed a general framework for analysing
the statistics of liquid structure and connecting explicitly these
correlations to the configurational entropy of the liquid. We have
argued that the fluctuation statistics of the high temperature liquid
(or, equivalently, a version of the liquid with only repulsions
retained) contain important information on the geometrical constraints
of the liquid, even in the supercooled regime. This proposal runs
counter to the current practice in computational studies of liquid
structure to look in the lowest temperature liquids, an approach
ideally suited to identify \emph{what} structures are present but
unhelpful in addressing \emph{why} these structures are so favoured.
We have addressed this \emph{why} question with an explicit expression
for the configurational entropy in terms of the statistics of the
local structures, revealing in particular the importance of the
covariance of the structural fluctuations.  With our new approach to
the analysis of liquid structure, we have thus presented in this
article a first step towards a structural characterization of
amorphous configurations that goes beyond merely descriptive
approaches.

As a testing ground for our theory, we have considered a large sample
of model liquids within our Favoured Local Structure lattice
model. Here, each liquid we consider is characterised by two FLS’s
drawn from a selection of low-symmetry structures. A modest
correlation between glass forming ability and the high $T$ covariance
$C_{ij}$ serves both to support our claims for the relevance of the
high temperature structural statistics. Beyond providing a glimpse on
the mechanisms leading to a glass transition, we note that this model
also gives us a reliable way to generate glass-forming systems in
discrete lattice models without spatial heterogeneity or frozen
disorder.  Such a simple geometrical model may prove useful to address
the challenges that remain in explaining the physics of the glass
transition.

\section{Acknowledgements}

{\bf Acknowledgements} PH acknowledges the financial support of the
Australian Research Council. PR is supported by ``Initiative Doctorale
Interdisciplinaire 2013'' from IDEX Paris-Saclay.

\end{document}